\documentstyle[twocolumn,aps]{revtex}
\begin{document} 
\title{
\begin{flushright}
\small
TIFR/CM/98/202(I)
\end{flushright}
Reentrant Peak Effect in an anisotropic
superconductor 2H-NbSe$_2$: Role of disorder\\
\vskip 0.25truecm
\normalsize
S. S. Banerjee$^{1,@}$, N. G. Patil$^1$,  S.
Ramakrishnan$^1$, A.K.
Grover$^1$ , S. Bhattacharya$^{1,3,*}$, P.K. Mishra$^2$, G.
Ravikumar$^2$,  T.V.
Chandrasekhar Rao$^2$, V.C. Sahni$^2$, M. J.
Higgins$^3$, C. V.
Tomy$^{4,+}$, G. Balakrishnan$^4$, D. Mck.
Paul$^4$. 
\vskip 0.1truecm
\noindent
{\it $^1$Tata Institute of Fundamental Research,
Mumbai-400005, India\\
$^2$TPPED, Bhabha Atomic Research Center,
Mumbai-400085, India\\
$^3$ NEC Research Institute, 4 Independence
Way,
Princeton, New Jersey 08540, U.S.A.\\
$^4$ Department of Physics, University of
Warwick, Coventry, CV4 7AL,
U.K.\\}
\bigskip
\small \vskip 0pt
\noindent \hfill 
The reentrant nature of the {\bf Peak Effect (PE)} is established in a single crystal of anisotropic superconductor 2H-NbSe$_2$ via electrical transport and dc magnetization
studies. The role of disorder on the reentrant branch of the {\bf PE} has been
examined in three single crystals with varying levels of quenched random disorder. Increasing disorder presumably shrinks the (H,T) parameter space over which
vortex array retains spatial order. Although, the upper branch of the {\bf PE} curve
is somewhat robust, the lower reentrant branch of the same curve is strongly affected by disorder.
\small
\begin{flushleft}
PACS numbers : 74.60-w - Type II Superconductivity.\\
PACS numbers : 74.25Dw - Superconducting phase diagrams.\\
PACS numbers : 64.70Dv - Solid-liquid transitions.\\
\end{flushleft}
\small
$^*$ Present and Permanent address: NEC
Research Institute, 4 Independence Way,
Princeton, New Jersey 08540, U.S.A.\\
$^+$ Present address: Department of Physics,
Indian Institute of Technology, Kanpur,
208016, India.\\
$^{@}$ Corresponding author, e-mail : satya@tifrc2.tifr.res.in \\
Keywords : Reentrance, Peak Effect curve, Effect of disorder, 2H-NbSe$_2$
}
\maketitle
\normalsize
\noindent
A magnetic flux line lattice (FLL) or vortex lattice
(VL) with lattice
constant $a_0$$\propto$$ {1 \over \sqrt Field}$
was predicted \cite{r1} 
in the entire mixed state of a 
type-II superconductor within a mean field model.
The advent of high 
temperature superconductors (HTSC) focused
attention on the 
influence of thermal fluctuations on VL and led to the 
prediction \cite{r2}
of a Vortex Liquid State with an unusual reentrant
phase boundary in the
magnetic field-temperature (H, T) plane such that
at a fixed T, the melting
phase boundary (H$_m$(T)) is encountered
{\it twice} upon increasing H \cite{r3}. A dilute vortex liquid phase is expected to
form above the lower critical field H$_{c1}$, where
a$_0$ $\geq$ $\lambda$, the
magnetic penetration depth $\lambda$ gives a measure of 
the range of repulsive interaction that governs the FLL. 
A highly dense vortex liquid phase is expected below the upper
critical field, where a$_0$ $\geq$ 2$\xi$; $\xi$ is the
coherence length, i.e., the radius of the
vortex core. The two branches comprising reentrant melting phase boundary
join around the so-called ``nose" temperature above which the 
vortex solid phase is thermally melted at all field values. The upper
branch of the melting curve is given approximately by eqn. 5.5 of Ref.3,
$$ B_m~=~\beta_m ({c_L^4 \over G_i}) H_{c2}(0) (1-{T \over T_c})^2, \eqno(1) $$
where c$_L$ is the Lindemann parameter, G$_i$ is the Ginzburg
number ($=$ (1/2)(k$_B$T$_c$ / H$_c$$^2$$\xi$$^3$$\epsilon$/8$\pi$)$^{1/2}$) and $\beta_m$ $\approx$ 5.6 . G$_i$ is much larger in the Cuprate HTSC systems as compared to that in low T$_c$ alloys, which facilitates the observation of the 
dense vortex liquid phase in HTSC systems \cite{r3}. 
However an experimental observation of the dilute vortex liquid phase, 
has remained elusive. On the lower branch, the melting phase boundary is governed by eqn. 5.19 of Ref.3,
$$ {\lambda \over a_0} e^{({a_0 \over \lambda})} = \beta_m ^\prime ({c_L^4 \kappa~ln \kappa \over G_i}) ({T_c \over T})^2 (1 - {T \over T_c}) \eqno(2) $$ where~ a$_0$$=$[2/3$^{1/2}$($\phi$$_0$/B)]$^{1/2}$~ and~ $\beta$$_m$$^{\prime}$$=$0.5. Solving eqn.(2) gives B$_m$(T$\rightarrow$0)~$=$~H$_{c1}$(0)[ln(T/T$_c$)]$^{-2}$. The large value of $\kappa$ in the Cuprates presumably restricts the dilute vortex liquid phase to very small induction values, making its experimental observation a challenging task.

A further difficulty in the observation of the vortex liquid phases arises due to the ubiquitous quenched disorder (i.e., pinning centers) in all real systems which adversely affects the stability of the ordered vortex solid phase. The effect of disorder is expected to be especially strong for dilute vortex arrays and may mask the observation of an underlying dilute liquid - dilute solid melting transition \cite{r4,r5}.

In the context of quenched disorder, the phenomenon of the peak effect (PE), an anomalous increase in the critical current, is continuing to receive a great deal of attention as it marks the competition between interaction and disorder (i.e., elastic energy versus pinning energy) \cite{r6,r7,r8,r9,r10,r11,r12,r13}. In a collective pinning description \cite{r14} of
superconductors,
J$_c$ relates inversely to the volume V$_c$ of
Larkin domain 
($J_c$ $\propto$ ${ 1 \over \sqrt V_c}$) over
which FLL is correlated.
The anomalous peaking in J$_c$ indicates
a rapid decrease in V$_c$ due to softening of elastic
moduli of FLL at the 
incipient melting transition
\cite{r6,r11,r12,r14}, thus making the PE yet another signature of the loss of order of the vortex array. The identification of PE as a ``phase boundary'' remains among the more complex and controversial questions in type II superconductors in general \cite{r6,r7,r8,r9,r10,r11,r12}. 

Ghosh {\it et al} \cite{r10}
showed that in a weakly pinned vortex lattice
of hexagonal 2H-NbSe$_2$
(T$_c$$\approx$7.1~K), the {\bf PE} 
curve, which is the locus of peak temperatures
T$_p$(H) (determined from the magnetic 
shielding response, i.e., ac
susceptibility $\chi^{\prime}$ measurements),
bore a striking resemblance to the theoretically proposed reentrant
\cite{r2,r3} melting phase boundary T$_m$(H).
The {\it turnaround} of reentrant T$_p$(H) curve
has been reported \cite{r10}
to occur when $a_0$(of 4000 A$^0$ at H $\approx$ 150 Oe) $\approx$$\lambda$$_c$ (in 2H-NbSe$_2$) , and it is
followed by a rapid broadening of 
{\bf PE}. The {\bf PE}
ultimately became undetectable
below 30 Oe \cite{r10}. Since PE is directly related to pinning, its variation with varying
pinning strength is crucial in providing an understanding of the complex interplay between 
thermal and quenched disorder in destabilizing the ordered phase. In this letter, we report new
results on the reentrant nature of
PE curve and the effect 
of quenched random disorder on it via electrical
transport measurements
and an equilibrium dc magnetization experiment
on a specimen of 2H-NbSe$_2$
utilized by Ghosh {\it et al} \cite{r10} and
magnetic shielding
($\chi^{\prime}$) response studies on two
different crystals of the same
compound with qualitatively different levels of
quenched random disorder. 
The conventional dc electrical transport results
not only fortify the
earlier claim \cite{r10} of reentrant nature of the
T$_p$(H) curve but also
corroborate the previous finding that {\bf PE}
broadens across fields where
{\it turnaround} in T$_p$(H) takes place.  The
magnetization hysteresis
data also confirm the double crossover of
T$_p$(H) curve in an isothermal
scan near the {\it turnaround} region.
We find that though the upper (higher H)
branch of T$_p$(H) curve
is somewhat robust, the lower reentrant branch is
strongly influenced
by disorder. Decreasing pinning strength results in the `nose'
region being pushed down to
fields lower than that reported earlier \cite{r10},
whereas increasing
pinning  could make the `nose' feature completely
disappear. These 
findings provide quantitative estimate on how increasing disorder shrinks
the (H,T) space
over which vortex array retains spatial order. 

2H-NbSe$_2$ has an appreciable value
($\approx$3 x 10$^{-4}$) of $Gi$, which lies
between
those of HTSC and conventional superconducting
alloys \cite{r11}. The weak
pinning characteristic of single crystals of this
compound 
[J$_c$/J$_0$$\sim$10$^{-5}$ to 10$^{-7}$,
where J$_c$ is the critical current 
density and J$_0$ ($\approx$ 10$^{8}$
A/cm$^2$) is the theoretical depairing current
density] facilitates
the formation of VL in them \cite{r11} and easy
variability of spatial
correlation lengths in its specimen make them
convenient test beds
to observe phenomena which are a consequence of
competition and interplay
between interaction, thermal fluctuations and
quenched random disorder.
Three single crystals with increasing quenched
disorder used in the present
study comprise platelets belonging to the same
batches as the specimen {\bf X}
used by Higgins and Bhattacharya \cite{r11}, as
the specimen {\bf Y}
used by Ghosh {\it et al} \cite{r10} and as the
specimen {\bf Z} used by
Henderson {\it et al} \cite{r13}. The dc resistance
data to ascertain 
{\bf PE} temperatures in specimen {\bf Y} were
recorded following
Ref. 11, the isothermal magnetization hysteresis
data to identify the
manifestation of {\bf PE} in a similar sample were
obtained on a
Quantum Design Squid Inc. magnetometer following
{\it the half scan technique} of 
Ravikumar {\it et al} \cite{r15} and ac
susceptibility data in all three
samples were recorded using a high sensitivity ac
susceptometer \cite{r16}.
The essential new findings are summarized in
Figs. 1 to 3. All data presented
are for H$\parallel$c, however, {\bf PE} is
observed in all orientations
in 2H-NbSe$_2$ and the anisotropic
Ginzburg-Landau description applies
to this system \cite{r11}.

Fig.1 shows the temperature variation of resistance, R vs reduced temperature (t$=$T/T$_c$(0)), in the crystal {\bf Y} at some of the low fields (120 Oe to 380 Oe) to locate the signature of
the reentrant behavior in T$_p$(H) data. The inset (a) of Fig.1 shows that the peak in J$_c$ at a given H manifests as an anomalous dip in R(t). As emphasized in Ref. 11, the location of each T$_p$(H) is robust. The main panel of Fig. 1 shows portions of R(t) curves at different H and focusses attention onto the identification of t$_p$(H) values. The inset panel (b) of Fig.1 depicts the PE curve which passes through the t$_p$(H) values for 120$\leq$H$\leq$380 Oe. The
shape of the PE curve clearly demonstrates its reentrant characteristics, 
with {\it turnaround} occurring at about 180 Oe, in excellent agreement with earlier results \cite{r10} obtained via $\chi^{\prime}$(T) measurements. The {\bf PE} behavior in
a given R(t) curve starts at a temperature, T$_{pl}$, where the resistance starts to decrease
(see t$_{pl}$ mark for H$=$360 Oe curve). The data in the main panel of Fig. 1 imply that
significant broadening of {\bf PE} occurs as H values are decreased below 200 Oe, and eventually below 120 Oe, the {\bf PE} becomes so shallow that t$_p$(H) cannot be located precisely for H$<$100 Oe.

A further confirmation of the reentrant characteristic of {\bf PE} curve (in sample {\bf Y} with T$_c$(0)$=$7.17 K) is provided by the results of dc magnetization hysteresis measurements, an example of which is shown in Fig. 2. In an isothermal scan at T$=$6.95 K, we expect to
encounter both the lower and upper branches of the reentrant {\bf PE} curve.  The inset panel of Fig. 2 shows a portion of the M-H hysteresis data at T$=$6.95~K. The {\bf PE} on the upper branch manifests as an anomalous opening of the hysteresis loop before reaching the H$_{c_2}$ value. According to Bean's Critical State Model\cite{r17}, magnetization hysteresis $\Delta$M(H)
(difference between forward and reverse magnetization values at given H) is a measure of J$_c$(H). To locate the existence of {\bf PE} on the lower branch and to elucidate that the peak field H$_p$ on the lower branch of {\bf PE} curve {\bf increases} as T {\bf increases}, we show in Fig. 2 the variation of 4$\pi$$\Delta$M ($\propto$J$_c$(H)) vs H at 6.9~K, 6.95~K and 7.0~K in the low field region. It is to be noted that for H$<$ 60 Oe, $\Delta$M(H) values at 6.95~K are {\bf lower} than those at 6.9~K and this reflects the normal behavior in J$_c$(H,T). However, a crossover in $\Delta$M(H) curves occurs at
about 70 Oe such that $\Delta$M(H) values at 6.95~K {\bf exceed} those at 6.9~K, thereby
exemplifying the observation of anomalous behavior \cite{r18} in J$_c$(H,T) due to the presence
of {\bf PE} on the lower branch at 6.95~K. It may be further noted that $\Delta$M(H) values at 7.0~K become larger than those at 6.95~K above 130 Oe. Thus, at 7.0~K the (lower) peak effect is seen to be encountered at a field value larger than that at 6.95~K, this establishes the
central characteristic (i.e., dT$_p$/dH$>$0) of the lower branch of the PE curve.

Fig. 3 shows the temperature
variation of the in-phase
ac susceptibility ($\chi^{\prime}$ vs reduced
temperature $t$) at a
frequency of 211 Hz and in h$_{ac}$ of 1
Oe(r.m.s.) at some of the chosen dc fields H$_{dc}$ (0 to 2 kOe) in three crystals {\bf X}, {\bf
Y} and {\bf Z} of 2H-NbSe$_2$.
These crystals were grown by the same vapor transport
method \cite{r19},
but with starting materials of different purity such
that the specimen
{\bf X} is the cleanest among the three and the
specimen {\bf Z},
the most disordered. The relative purity of the three crystals
is reflected
in the {\it width} of the superconducting transition in
$\chi^{\prime}$(t)
response in zero field, which can be seen to
progressively enhance in
Figs. 3(a) to 3(c). In a
$\chi^{\prime}$(t) measurement at a given H,
the {\bf PE} is identified
by an anomalous negative peak \cite{r9,r10}. The
peak temperatures T$_p$(H) are independent of 
frequency and amplitude of ac
field h$_{ac}$ in the parametric range of our experiment \cite{r20}. Fig.
3(a) shows that in the most 
weakly pinned crystal {\bf X}, the {\bf PE} peaks
are very sharp;
{\it their half widths are smaller than the width of
superconducting transition
in zero field}. This fact supports the
association of T$_p$(H) with a possible phase transition,
like, the phenomenon of change in spatial order of
the vortex
array across T$_p$. 

Fig. 3(b) is based on the $\chi^{\prime}$(t)
response in crystal {\bf Y} \cite{r10}. 
T$_p$(H) values for H$=$300 Oe and H$=$100
Oe can be seen to be nearly the same, consistent
with the reentrant nature of {\bf PE} curve in
crystal {\bf Y} \cite{r10}. However, the
{\bf PE} in H$=$200 Oe (data not shown) 
is significantly broader
than that at H$=$300 Oe. 
Fig. 3(c) shows the $\chi^{\prime}$(t) response
in the most strongly pinned
\cite{r13} crystal {\bf Z}. In this sample, the
{\bf PE} peak is
considerably broad even at H$=$1000 Oe as
compared to the corresponding peaks
in crystals {\bf X} and {\bf Y} (cf. Figs. 3(a) to
3(c)). The {\bf PE} peak
is barely discernible at H$=$300 Oe in crystal {\bf
Z} and below 200 Oe, the
{\bf PE} manifests only as a hump/kink in
$\chi^{\prime}$(t) curve.

Fig.4 summarizes the T$_p$(H) data in three
crystals {\bf X}, {\bf Y} and
{\bf Z} as H vs T$_p$/T$_c$(0) curves. For the
sake of convenience and 
reference, we have also included in
this figure the H$_{c_1}$(t)
and H$_{c2}$(t) curves
obtained in crystal {\bf X} \cite{r6}. Thus, from
Fig.3 and Fig.4, it is
apparent that enhanced quenched random
disorder (pinning) results in the
following effects :\\
(i) For a given H (H$\geq$300 Oe), the {\bf PE}
peak occurs at a lower value 
of t (see, for instance, arrows marking T$_p$(H)
values in three crystals at H=300 Oe in Figs.3(a) to 3(c)). This amounts
to a ``lowering'' of the upper 
branch of {\bf PE} curve with enhanced pinning (see Fig.4). Also note that the {\bf PE} broadens as the disorder
increases.\\ 
(ii) The reentrant lower branch of {\bf PE} curve
is clearly evident
only for crystal {\bf Y} with intermediate level of
disorder and not so for
the other two crystals. The {\bf PE} curve in crystal {\bf X} shows a steep fall for 200 $<$ H $<$ 30 Oe and its turnaround characteristic presumably lies much below 30 Oe; however, {\bf PE} in $\chi^{\prime}$(T) data is unobservable in crystal {\bf X} for H$\leq$20 Oe. In sample {\bf Z}, the {\bf PE} becomes so broad
at 300 Oe that the precise
location of T$_p$(H) below this field
becomes somewhat ambiguous (see Fig.3(c)). However, 
it can be safely surmised that {\bf PE} curve at
lower fields (H$<$ 500 Oe)
moves away from the values that can be extrapolated from
T$_p$(H) vs H data at higher
fields (H$>$500 Oe). Thus the effect of enhanced disorder ({\bf X} to {\bf Z}) is seen to result in ``raising'' to higher fields, the lower reentrant branch of the PE curve (in contrast to the upper branch).

The results of Figs. 3 and 4 provide the following general conclusions:\\
(i) For the upper branch (i.e., dT$_p$/dH$<$0) of the PE curve, increasing pinning reduces the 
volume of Larkin domain of FLL, which then requires less thermal fluctuations to melt/amorphize it around t$_p$. The quenched disorder and thermal fluctuations conspire together to destabilize
the ordered phase and a ``lowering'' of the t$_p$(H) curve is seen.\\
(ii) For the lower branch (i.e., dT$_p$/dH$>$0), increasing pinning reduces the volume of Larkin domain as before. Thus, one 
needs enhanced interaction (i.e., increase in H) to stabilize the ordered phase and the ``raising'' of the PE curve occurs.\\
(iii) Increasing pinning has the general effect of masking the difference between a vortex solid and a vortex liquid, converting both into a ``glassy'' state. A clear distinction between the two phases is no longer possible, the transformation process is more gradual and thus a broadening of the PE results.\\
(iv) As can be ascertained from the main panel of Fig.1, the onset (at t$_{pl}$) of PE, which is apparently driven by quenched disorder, is also reentrant, just as the vortex melting transition is expected to be. Recent muon spin rotation studies \cite{r21} on crystals of 2H-NbSe$_2$ have provided microscopic evidence in favour of a sudden change in spatial order of FLL at the onset of PE, i.e., at T$_{pl}$.  
 
Finally, it is of interest to examine the scenario
emerging from Fig.4
with the expectations of theoretical studies in the
(H, T) region of our experiments. In the cleanest
crystal {\bf X}, where
we observe only the upper branch of {\bf PE}
curve, which most closely marks the transformation of the ordered solid into the {\it pinned ``liquid'' state} and a narrow region 
separates this line from T$_c$(H)
line (cf. Figs. 6, 24 and 25 of Ref. 3 and our
Fig.4). In between T$_p$(H)/H$_p$(T)
and T$_c$(H)/H$_{c_2}$(T) lines, the pinned
liquid becomes unpinned
at H$_{irr}$(T) (see Fig. 2(a) for location of
H$_{irr}$ at 6.95~K). 
Fitting the higher field (H$>$200 Oe) branch of {\bf
PE} curve to a standard
quadratic Lindemann relationship (see eqn. (1)) 
with Gi$\sim$3$\times$10$^{-4}$ and
H$_{c_2}$(0)$\approx$4.6~T \cite{r6}, we get
a reasonable value for Lindemann number
c$_L$$=$0.15. A pronounced
departure (in T$_p$(H) at H$\leq$200 Oe) away
from the upper portion
appears in accord with a recent
theoretical work of Blatter
and Geshkenbein
\cite{r22}, who found a similar rapid drop in
FLL melting curve (in the absence of pinning effects),
 away from
the H$_{c_2}$ phase boundary, at low fields (i.e.,
above the 
{\it turnaround} feature of the melting boundary).
The reentrant lower
branch of {\bf PE} curve may thus be at still
lower fields in crystal
{\bf X}, as proposed theoretically \cite{r22}, and
outside the detection capability
of the present measurements. In crystal {\bf Y}, with
intermediate level of
disorder, Ghosh {\it et al} \cite{r10} estimated
c$_L$ as about 0.17 from
higher H portion of the T$_p$(H) data. They also
showed that if the 
Nelson-LeDoussal line \cite{r4} indeed marks the
crossover from an interaction dominated
to a disorder-dominated region, then the ratio of
entanglement length
L$_E$ to the pinning length L$_c$ becomes
approximately equal to 1 
(L$_c$/L$_E$$\sim$1) at B$=$30 Oe and 
t$=$0.975. L$_c$/L$_E$ is given as \cite{r3}, 
$$ {L_c \over L_E}~=~({ \pi \kappa^2 ln(\kappa)
\over \sqrt 2})~ 
({a_0^2 \over 2 \pi \lambda (0)})~{(J_c)^{1/2} \over
(G_ij_o)^{1/2}} ~
{(1-t)^{4/3} \over t}, \eqno(1) $$
where various 
symbols have their usual meaning (see eqn. 6.47 of Ref.3).
Below 30 Oe, the vortex array 
would be in disentangled liquid or glass state
\cite{r3,r4,r5}. Eqn.1 implies that the field 
at which
crossover to glassy state
can occur is $\propto$ J$_c$$^{1/2}$. J$_c$ is 
larger in
crystal {\bf Z} by nearly a factor of 50
as compared to that in crystal {\bf Y} .
This, coupled with the observation that in crystal
{\bf Z} the {\bf PE} 
becomes difficult to discern below
t$\approx$0.96, yields a value of crossover 
field of $\sim$400 Oe in sample {\bf Z}, which
demonstrates a very reasonable agreement
with an observed value of 300 Oe in our experiment. 

In summary, we have demonstrated the variation of the PE at high temperatures
and low fields, and especially of its lower reentrant, branch is experimentally measurable.
The results show not only the shifts in the relevant transitions/crossover lines separating the ordered phase from the disordered ones, they also illustrate how the distinction between them is blurred as quenched disorder is varied. Some aspects of the results could be semi-quantitatively explained by the available theories \cite{r4,r5,r22,r23}. Nevertheless, a more detailed theoretical analysis of the low field - high temperature regime, especially on the liquid - dilute solid transformation process, is necessary. \\
We thank Prof. R. Srinivasan for a critical reading of this manuscript.

\begin{figure}
\caption{ The inset panel (a) shows the temperature 
variation of Resistance (R vs reduced temperature, 
t=T/T$_c$(0)) in H$_{dc}$$=$280 Oe ($\parallel$c) in the crystal {\bf Y}
of 2H-NbSe$_2$. The Peak Effect (PE) region can be identified by an
anomalous dip and peak temperature t$_p$ is marked. The main panel
shows portions of R vs t curves in the PE regions at some chosen dc
fields in the range 140 Oe $\leq$ H $\leq$ 360 Oe. The marked
t$_p$(H) values elucidate that t$_p$(H) increases as H decreases 
(dH/dt$_p$ $<$ 0) from 360 Oe to 200 Oe and thereafter (for H$<$ 200 Oe)
dH/dt$_p$ $>$ 0. The inset panel (b) focusses attention onto the
reentrant nature of locus of t$_p$(H).}
\label{Fig.1}
\end{figure}
\begin{figure}
\caption{Isothermal magnetization hysterisis data 
[4$\pi$$\Delta$M $=$ 4 $\pi$ (M$_R$-M$_F$)] for 
H$_{dc}$$\parallel$c in the crystal Y of 2H-NbSe$_2$ at the temperatures
indicated. The arrows identify the fields corresponding to onset
of PE phenomenon on the lower (reentrant)
branch of PE curve at temperatures of 6.95 K and 7.0 K. It is
to be noted that the field at which (lower) PE occurs at 7.0 K
is larger than that at 6.95 K. The inset panel shows a portion
of the M-H loop showing upper PE at 6.95 K. (Upper) Peak field
H$_p$ and irreversibility field H$_{irr}$ at 6.95 K have 
been identified.}
\label{Fig.2}
\end{figure}
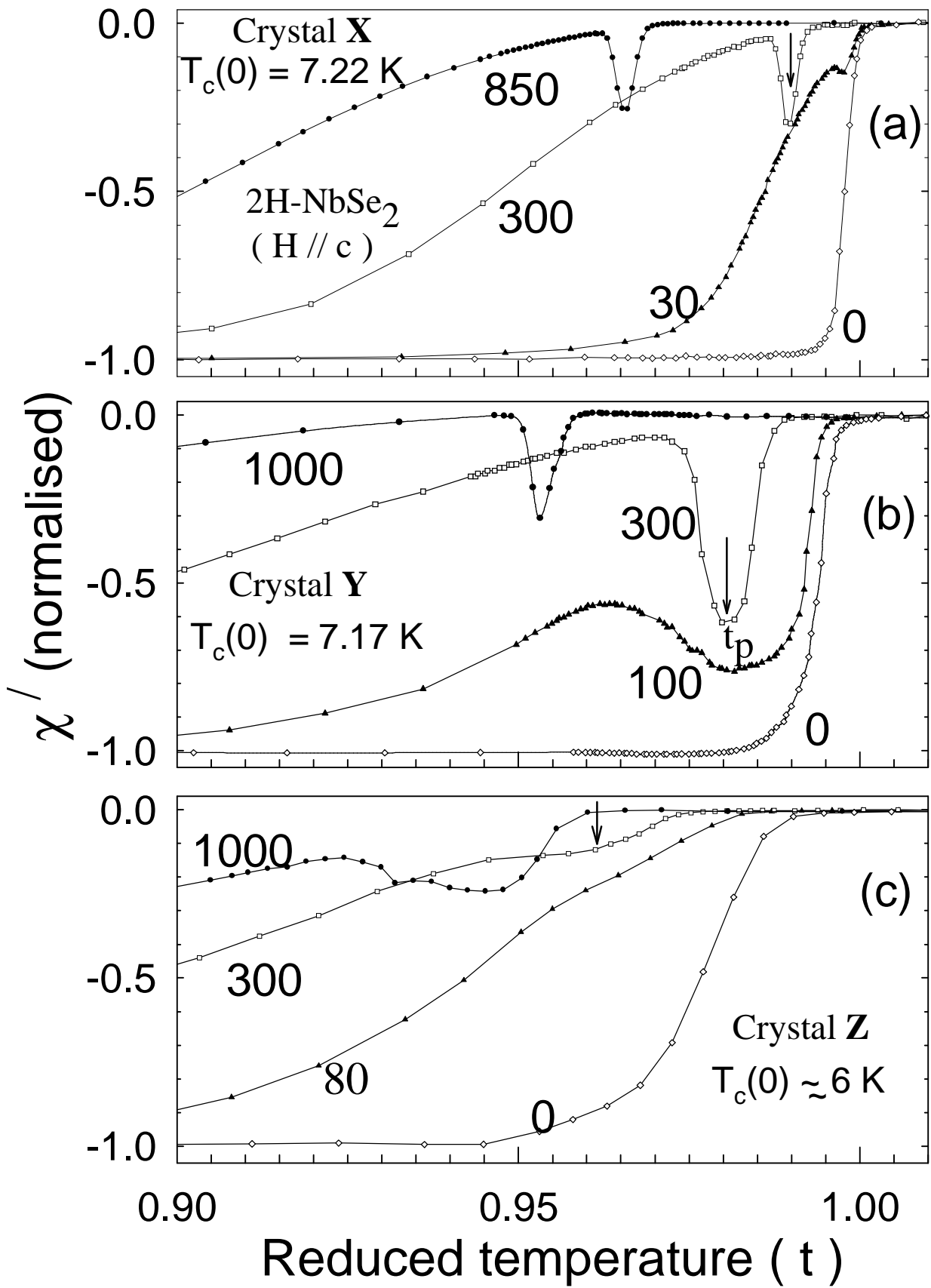
\begin{figure}
\caption{ In phase AC (f=211 Hz, h$_{ac}$ $=$ 1 Oe (r.m.s.)) susceptibility
($\chi$$^{\prime}$) vs reduced temperature (T/T$_c$(0)) in crystals
{\bf X}, {\bf Y} and {\bf Z} of 2H-NbSe$_2$ in fixed dc fields 
($\parallel$ c) as indicated. The arrows mark the t$_p$ values at
H$_{dc}$$=$300 Oe in different crystals.}
\label{Fig.3}
\end{figure}
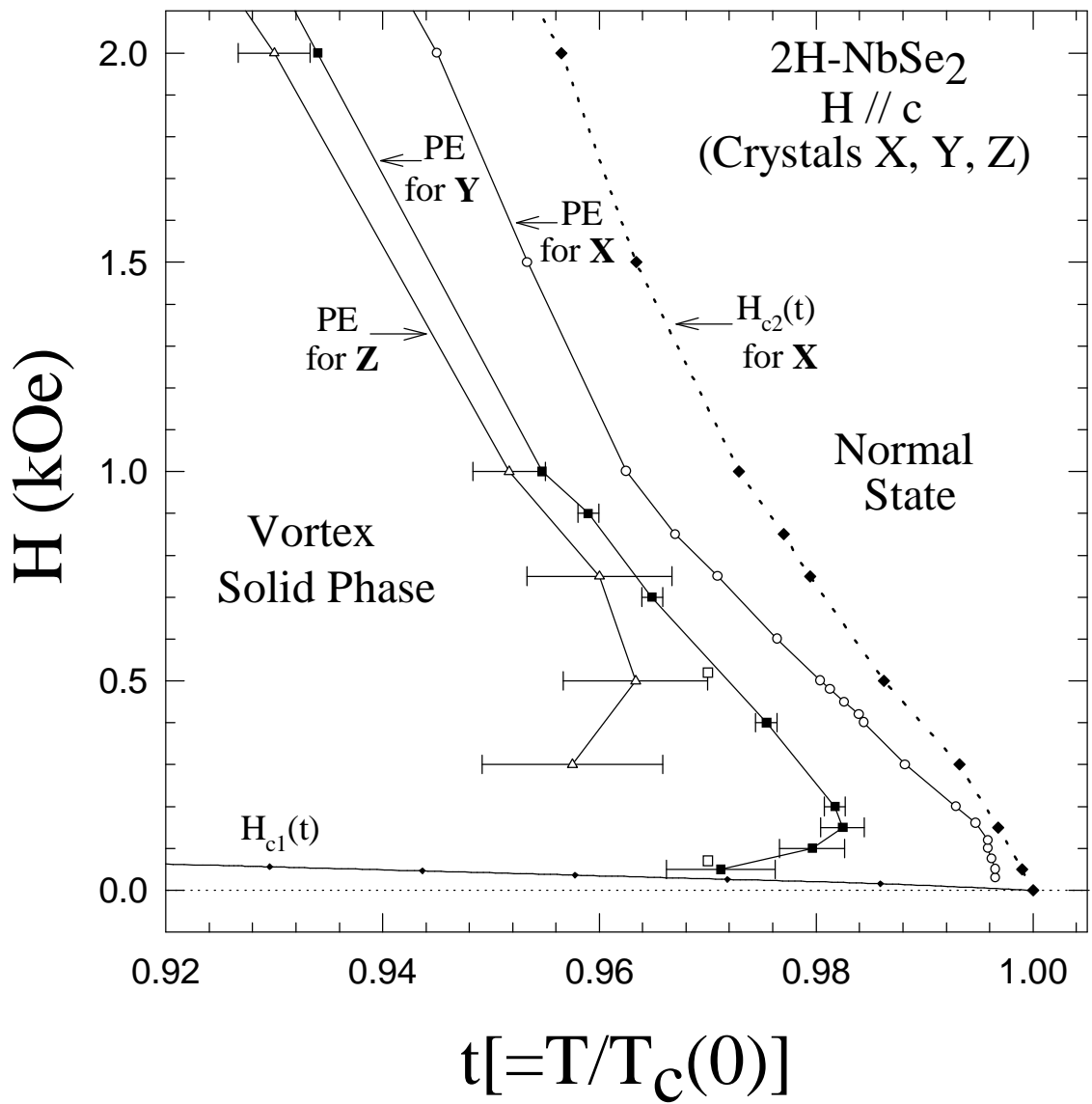
\begin{figure}
\caption{Magnetic phase diagram in three crystals of 2H-NbSe$_2$. 
PE curves and H$_{c2}$(T) curve correspond to T$_p$(H)/T$_c$(0)
and T$_c$(H)/T$_c$(0) values obtained from $\chi$$^{\prime}$(t)
data as in Fig.3. For crystal Y, the two open squares lying on the
upper and lower branches of PE curve identify the peak field (H$_p$)
and crossover field at 6.95 K as in Fig.2.}
\label{Fig.4}
\end{figure}
\end{document}